# Luminescent Trityl-based Diradicaloids: A Theoretical and Experimental Assessment of Charge-Resonance in Low-Lying Excited States

Davide Mesto,[a,‡] Michele Orza,[b,‡] Brunella Bardi,[c] Angela Punzi,[a] Imma Ratera,[d] Jaume Veciana,[d] Gianluca Farinola,[a] Anna Painelli,[c] Francesca Terenziani*,[c] Davide Blasi*,[a] Fabrizia Negri*,[b,e,f]

a) *Dipartimento di Chimica, Università degli Studi di Bari Aldo Moro, 70125 – Bari, Italy*
b) *Dipartimento di Chimica "Giacomo Ciamician", Università di Bologna, 40129 – Bologna, Italy*
c) *Dipartimento di Scienze Chimiche, della Vita e della Sostenibilità Ambientale, Università di Parma, Parco Area delle Scienze 17/a, 43124 Parma, Italy*
d) *Institut de Ciencia de Materials de Barcelona (CSIC)/CIBER-BBN, Campus de la UAB, 08193 -Bellaterra, Barcelona, Spain.*
e) *INSTM UdR Bologna, 40129 Bologna, Italy.*
f) *Center for Chemical Catalysis—C3, Università di Bologna, 40129 Bologna, Italy*

‡D. M. and M. O. contributed equally to this manuscript.

**ABSTRACT:** The tris(2,4,6-trichlorophenyl)methyl radical (**TTM**) has inspired the synthesis of several luminescent diradicals and diradicaloids, providing an extraordinary opportunity to control the nature of the low-lying excited states by fine-tuning the diradical character. However, the photophysical properties of **TTM-**derived diradicals remain not fully understood yet. Here we present a combined theoretical and experimental investigation on **TTM-**derived diradicals to elucidate the origin of their luminescence. The theoretical analysis focuses on a series of symmetric **TTM-**derived diradicals with singlet ground state, featuring radical moieties linked by π-conjugated spacers of different length. The nature of the lowest excited electronic states that control their photophysical behaviour is discussed in detail. The study is complemented by a complete spectroscopic characterization of the **TTM-TTM** diradical, synthesized using a novel, simpler and more efficient procedure exploiting the unique reactivity of **TTM**. The lowest excited states of the diradicals differ qualitatively from those of **TTM**: two novel low-lying states emerge in the diradical, due to charge resonance (CR) between the two radical units. The lowest CR state is a dark state for symmetric diradicals. The CR nature explains the blue-shifted emission observed by increasing the distance between the radical centres as seen in **TTM-ph-TTM**. This insight suggests different design strategies to improve the luminescence properties of **TTM-**derived diradicals.

## Introduction

Luminescent molecules are attractive functional materials for application across different fields, including light-emitting devices,[1–3] bio-imaging[4,5] and, more recently, quantum technology.[6] The tris(2,4,6-trichlorophenyl)methyl radical (**TTM** in **Figure 1a**) is an open-shell building block that has driven the molecular design of several doublet emitters, which have recently received great attention due to their unique photonic and electroluminescence properties.[7–11] More recently, **TTM** has inspired the synthesis of several diradical and diradicaloid species, with the potential to create a disruptive impact on various technological fields, including organic light-emitting diodes (OLEDs), magnetoluminescence and quantum information science, among others.[12–19] However, unlike the luminescent free radicals, the photophysics of diradicals derived from **TTM** remains not fully understood yet being essential for many new applications.

**Figure 1** shows the chemical structure of **TTM** and recently reported **TTM**-derived diradicals, giving schematic representation of their absorption and fluorescence spectra. **TTM** shows intense absorption bands in the 350-400 nm spectra region and two very weak absorption bands at 500 nm (ε = 900 cm$^{-1}$M$^{-1}$ in CHCl$_3$) and 545 nm (ε = 1000 cm$^{-1}$M$^{-1}$ in CHCl$_3$).[20] Such bands, named D bands by Ballester *et al.*[21], are due to the alternant symmetry of the molecule, determining the degeneracy of the HOMO→SOMO and SOMO→LUMO gaps. UV and visible excitations have similar transition dipole moments, and their mixing leads to an out-of-phase combination at low energy (D$_1$ state at 545 nm) and an in-phase combination at high energy (D$_2$ state at 370 nm)[22,23].

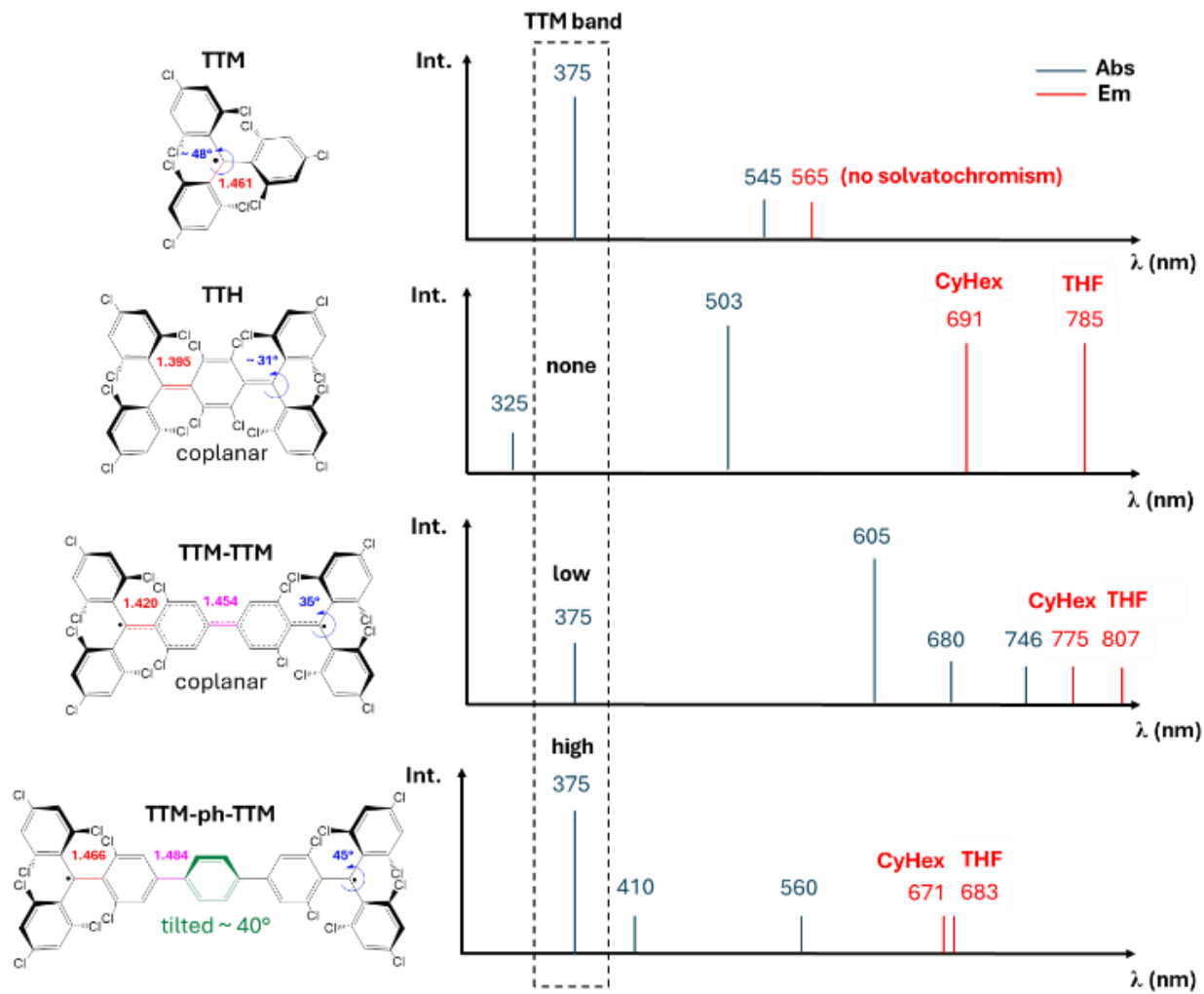

**Figure 1.** (a) Molecular structure, bond lengths, dihedral angle of torsion (determined by single crystal X-ray diffraction) of **TTM** and three diradicaloids based on **TTM**, namely **TTH**, **TTM-TTM** and **TTM-ph-TTM** (b) Schematic representation of the main bands in absorption and emission spectra. For the latter the solvatochromic effect is also indicated.

The $D_1$ state shows a weak oscillator strength, responsible for the low photoluminescence quantum yield (PLQY) of the **TTM** in solution (PLQY ≈ 1-2% in several solvents). Recent studies have rationalized the weak emission from this formally forbidden state as due to the mixing of intramolecular charge transfer (CT) excitations between the radical center and its ligands[24].

Moving to diradical species, in a previous work[12] some of us reported the synthesis and photophysical characterization of **TTH**, a **TTM**-like Thiele hydrocarbon, that shows a singlet ground state and moderate diradical character. This system displays a unique photophysical behavior, remarkably different from the parent **TTM** radical. Notably, it features an intense and red-shifted absorption at 500 nm (ε = 48200 cm$^{-1}$M$^{-1}$ in CHCl$_3$),[12] a large Stokes shift and remarkable solvatochromism, unexpected for a non-polar centrosymmetric molecule. Quantum-chemical calculations revealed that the emission arises from an unusual mechanism involving the mixing of two charge-resonance states (as in the sudden polarization mechanism[25] occurring in the excited ethylene). This mixing, combined with the twisting of the strongly elongated exocyclic bond in the excited state, leads to the formation of a charge-separated, broken-symmetry state. Although the broken-symmetry geometry clearly justifies the progressive red-shift of the **TTH** fluorescence band with increasing solvent polarity,[26] the mechanism responsible for the distorted geometry, supported also by time-resolved measurements and invoked for some perfluoro-Thiele hydrocarbons,[27] has never been reported for more conventional quadrupolar dyes.

Very recently, the Chichibabin hydrocarbon **TTM-TTM** has been reported[14,28] and its properties have been discussed by von Delius et al. **TTM-TTM** exhibits an intense absorption band at 605 nm and an emission at 780 nm, both strongly red-shifted compared to **TTM**. This phenomenon has been ascribed to the increased conjugation. However, the nature of the lowest lying excited states and their correlation with those of **TTH** and core-extended **TTM**-derived diradicals such as the Müller hydrocarbon **TTM-ph-TTM**[16], and **TTM** itself, remains unclear. As recently underlined by Abdurhaman et al.[28], **TTM-TTM** displays peculiar features in its absorption spectrum, resembling those of **TTM** but shifted to lower energies. In addition to its maximum absorption peak at 605 nm, two weak absorption bands are observed at 680 nm and 745 nm (**Figure 1**). Similar to **TTM**, in **TTM-TTM** the emission arises from the lowest energy state, according to Kasha's rule, with a PLQY ≈ 1%.

**TTM-ph-TTM**, has been recently reported to be a luminescent diradical[16] with an open-shell singlet ground state and a thermally accessible triplet state. Its UV/Vis absorption spectrum (in cyclohexane solution) exhibits three characteristic bands at 375 nm, 410 nm and 560 nm. The absorption band at 560 nm, although weak, is much stronger than the 545 nm absorption of **TTM**, and the emission (at 671 nm, PLQY = 0.4%) is significantly red-shifted compared to **TTM** but blue-shifted compared to **TTM-TTM**. This behaviour suggests that factors other than the extension of conjugation play a key role in determining the photophysics of the entire series of **TTM**-derived diradicals.

To overcome the low PLQY of **TTM**-based diradicaloids, it is urgent to gain a deeper understanding of their luminescence properties and the nature of emitting states. This is the main objective of this paper that will clarify whether the design principles established for **TTM** can be extended to these derivatives, enabling the development of more efficient luminescent materials.

In recent years, significant efforts were focused on designing stable conjugated open-shell diradicaloid molecules with a singlet ground state but variable diradical character[29–31], quantified by the diradical index $y_0$, ranging from 0 (closed-shell, CS) to 1 (pure open-shell, OS, perfect diradical).[32,33] The simplest reference molecular framework for understanding the electronic structure of these molecular systems is a CS molecule that progressively transforms into a diradicaloid and, ultimately, into a pure diradical. Alternatively, non-covalent, weakly interacting dimers of radicals can serve as a reference framework for the discussion of diradicals in which the radical centres are strongly localized and spatially distant. These two distinct reference frameworks have led to the development of two widely used approaches or models, each traditionally favored by different research communities.

In the first model, the focus is on a molecular system, and the low-lying electronic states of the diradicaloid/diradical molecule are generated by a configuration interaction involving two electrons in two orbitals (2e-2o).[34–38] Focusing either on the delocalized frontier HOMO (H) and LUMO (L) orbitals (Figure 2a), or on the localized orbitals A and B (Figure 2b), one triplet and three singlet states are generated. For homo-symmetric systems, the lowest energy singlet $S_0$ is characterized, in the CS limit, by the configuration $H^2$ (doubly occupied HOMO); for increasingly large diradical character, $H^2$ combines with the configuration $L^2$ (doubly occupied LUMO), so that, in the limit of the perfect diradical, $S_0$ is described by the *covalent* $H^2$-$L^2$ combination (Figure 2c). The other two singlets are characterized, respectively, by the H → L single excitation (HL-LH combination, SE) and by the (H,H → L,L) double excitation ($H^2+L^2$ combination, DE). SE is a dipole-allowed (bright) state, while DE is a dipole-forbidden (dark) state. SE and DE are charge resonance (CR) states resulting from the combination of two charge transfer (CT) *zwitterionic* configurations ($A^2$, $B^2$ in Figure 2b,c). As the system evolves from a CS molecule to a pure diradical, the CR component in the lowest singlet state is gradually reduced, by mixing with the $L^2$ configuration, rendering the state *covalent* (or *neutral*) in the perfect diradical limit. Conversely, the CR character in the DE state increases, by mixing with the $H^2$ configuration, ultimately reaching full CR character as indicated in the pure diradical limit in Figure 2c. The nature of the ground state (triplet or singlet) is determined by the combination of ferromagnetic coupling (FC) and antiferromagnetic coupling (AFC) contributions, and by the effects of spin polarization due to the other electrons in the system. Similarly, the order of the two singlet excited states SE and DE, is influenced and altered in real molecules, by electron correlation effects (Figure 2d). In previous works, some of us have shown that the distinctive photophysical properties of most conjugated diradicals are consistent with the DE state being lower in energy than the SE state.[39–44]

The Hubbard dimer model offers an alternative description, leading to valuable insights into the mechanisms of electron localization and its importance in materials' design.[45,46] In the Hubbard model, two radical centers, typically non-covalently bound, interact. It is directly comparable with the 2e-2o model in the localized orbital basis (Figure 2c), assuming that the exchange interaction $K_{A,B}$ between localized electrons is neglected. If the two radical sites are very far apart (perfect diradical) the hopping integral $t$ between the two site orbitals is negligible and the degenerate singlet and triplet states are *neutral* (*covalent* in the 2e-2o model). These states are separated by an energy $U_{eff} = U - V$ ($U$ is the on-site Coulomb repulsion, $V$ the inter-site Coulomb repulsion) from the two degenerate *zwitterionic* (or CR) states that essentially coincide with the SE and DE states of the 2e-2o model. For finite $t$, the diradical acquires a diradicaloid character, with the lowest singlet mixed up with the symmetric linear combination of the two zwitterionic configurations (*i.e.* the DE state), mirroring the increase of the $H^2$ contribution in the 2e-2o model, while the antisymmetric singlet (the SE state) stays unmixed. The Hubbard dimer model has been previously employed to identify the CR character of low-energy transitions in dimers of radicals, like e.g. (TTF$^+$)$_2$ or (TCNQ$^-$)$_2$ and to describe their low-energy photophysics.[47–49]

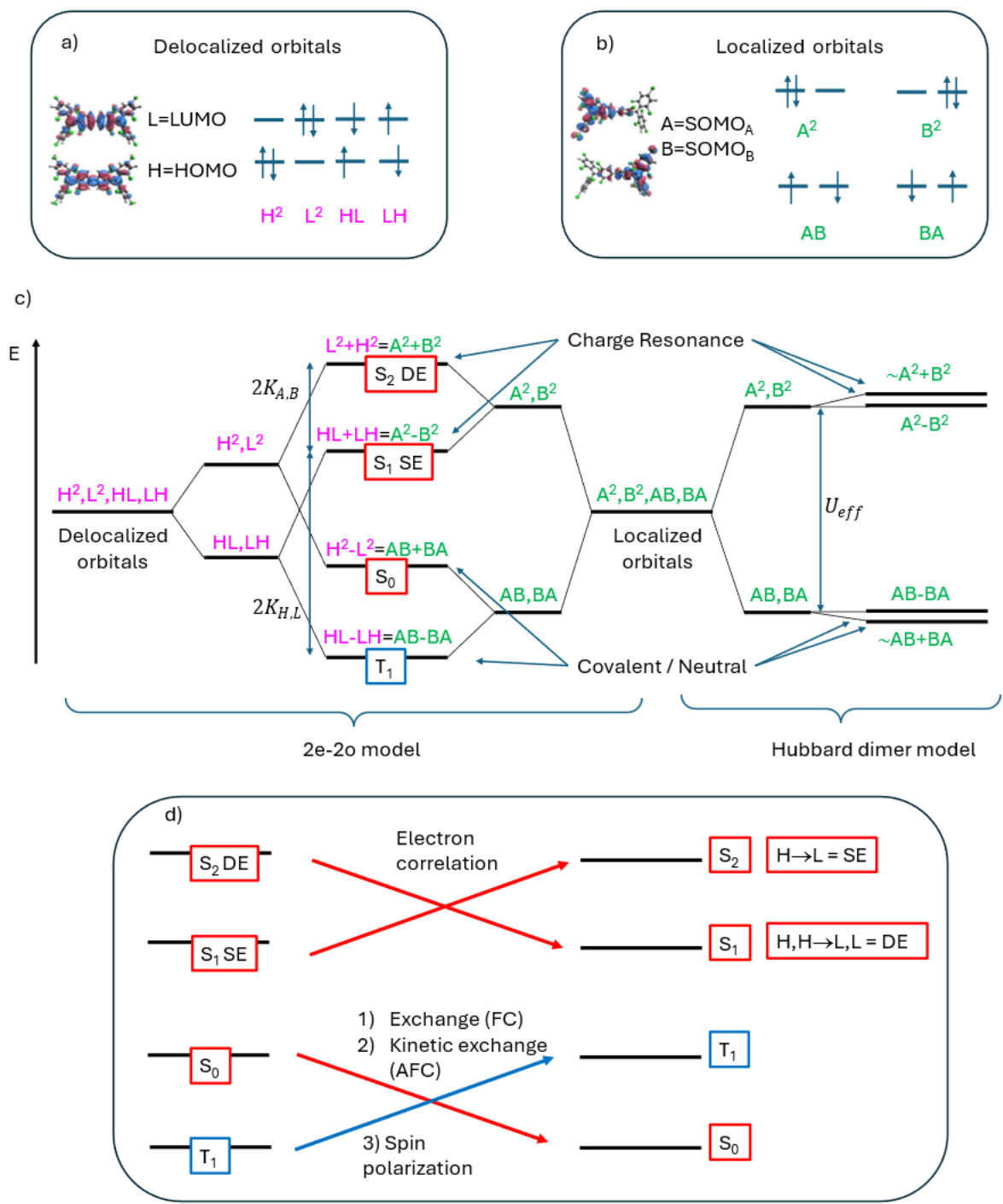

**Figure 2.** Comparison between the 2e-2o model and the Hubbard dimer model commonly employed to describe the electronic structure of diradicals. Two sets of molecular orbitals can be chosen for the 2e-2o model: delocalized or localized: (a) Electronic configurations in the delocalized orbital basis (magenta labels). The delocalized nature of the orbitals H and L is illustrated by the example on the left. In a perfect diradical these levels are degenerate. (b) Electronic configurations in the localized orbital basis (green labels). The A and B orbitals are the SOMO centered on the two radical moieties as shown by the example on the left. (c) Comparison between the two models. The spatial wavefunctions are indicated with magenta and green labels. For a perfect diradical the triplet and the lowest singlet states of the Hubbard model are degenerate. The small energy splitting shown on the right of panel c) corresponds to a diradicaloid rather than a perfect diradical. (d) Energy order of a real system: this panel shows that the energy order of the four states is affected by spin polarization and electron correlation, leading, in most diradicals, to a singlet ground state and a lowest lying DE excited state.

The states described by these two models are expected to govern the low-energy photophysics of **TTM**-derived diradicals, a family of systems that offer an unique opportunity to tune the diradical character and modulate the nature of the low-lying excited states. A thorough understanding and accurate classification of the excited states of **TTM**-based diradicals are essential steps for leveraging these intriguing species in practical applications.

Building on these concepts, we present a comprehensive theoretical analysis of a series of **TTM**-derived diradicals, specifically **TTM-TTM**, **TTM-ph-TTM** and **TTM-ph-ph-TTM**, characterized by an increasing distance between the radical centers, and singlet ground states. To investigate the nature of the electronic excited states that govern their photophysical behavior, a fragment orbital analysis is exploited to map the states of the diradicals in terms of those of their constituent mono-radicals. It will be shown that the nature of the lowest energy transitions in **TTM**-derived diradicals featuring a singlet ground state, involves novel low-lying excited states corresponding to the charge resonance between the two radical units, the lowest of which is a dark state. The computational analysis is validated by an extensive spectroscopic characterization of **TTM-TTM**, taking advantage of a novel, efficient synthetic route for this delicate diradical that exploits the unique reactivity of the **TTM** radical.

## Results and discussion

### Computational results

Seeking to find the parentage of diradical excited states to those of the **TTM** radical derivatives, we express each molecular orbital of the diradical in terms of their radical fragments. The fragment orbital interaction diagram resulting from such analysis is summarized in Figure 3 for **TTM-TTM** and in Figure S8 for **TTM-ph-ph-TTM**. The interaction diagram shows that both HOMO and LUMO delocalized orbitals (Figure 3b) of **TTM-TTM** correspond to linear combinations of the singly occupied molecular orbitals (SOMOs) of **TTM** fragments (see Table S4 for combination coefficients). Similarly, the localized orbitals from unrestricted DFT (UDFT) calculations (Figure 3c,d) not only correlate with the SOMOs of the **TTM** fragments but overlap almost entirely with the SOMO of one of the two fragments (Table S4). Accordingly, in Figures 2 and 3 these localized orbitals of the diradical are simply labelled SOMO$_A$ and SOMO$_B$. Figure 2 summarizes how the low-energy electronic states of diradicals can be effectively described by the 2e-2o model, adopting either a delocalized or a localized orbitals basis [35,36,39]. When localized orbitals are considered, the two lowest excited states (DE and SE) are dominated by the SOMO$_A^2$ ± SOMO$_B^2$ configuration (Figure 4). Since the ground state is described by the SOMO$_A$SOMO$_B$ configuration, the DE and SE states have a clear CR character, being composed[35,39,40,43] by the combination of the two charge transfer (CT) excitations, (SOMO$_A$→SOMO$_B$) ± (SOMO$_B$→SOMO$_A$) (Figure 4). Thus, considering the nature of the HOMO and LUMO orbitals in these **TTM**-derived diradicals, both 2e-2o and Hubbard dimer models align in identifying the lowest excited states as the DE and SE states.

A more realistic description of the lowest lying excited states of diradical systems requires to go beyond the simple 2e-2o or Hubbard models and, due to remarkable electron correlation effects, highly correlated multi-reference methods are required. Some of us demonstrated that expensive multi-reference methods can be replaced by cost-effective single-reference approaches based on time-dependent DFT calculations using an unrestricted reference configuration (TDUDFT), which in systems with a large diradical character properly capture the nature of the lowest excited states, included the elusive DE state[39,40,43]. **TTM-TTM**, **TTM-ph-TTM** and **TTM-ph-ph-TTM** display a singlet ground state with increasing $y_0(PUHF)$ values, from 0.93 to 0.99 (Table S5), which makes TDUB3LYP calculations suitable to explore their low-lying excited states. These calculations identify unambiguously the lowest singlet excited state with the DE state, as demonstrated by their dominant SOMO$_A^2$+SOMO$_B^2$ configuration (Figures S11-S13). The DE nature of the lowest excited state is further supported, for the three diradicals, by multi-reference CASSCF+NEVPT2 calculations (Table S6).

To summarize, both NEVPT2 and TDUB3LYP calculations identify the lowest excited state of **TTM-TTM** and related diradicals as the dipole forbidden, charge resonance DE state. The charge transfer nature of the DE and SE states is further confirmed by natural transition orbital (NTO) analyses (Figures S14-S16) and using TheoDORE[50] (Tables S7-S9 and Figures S17-S19). While the localized frontier orbitals of **TTM-TTM** are related to the SOMOs of the **TTM** fragments, the SE and DE states are *new* states, with a clear inter-radical charge-resonance, and they do not have counterpart in the isolated monoradical. The calculations show that both **TTM-TTM** and **TTM-ph-TTM** weakly emit from the DE state, while the SE state is responsible for the first intense absorption band (see Table S6 and Figure S20). Notably, moving from **TTM-TTM** to **TTM-ph-TTM**, the experimental absorption and the fluorescence features blue-shift, in striking contrast with considerations related to the conjugation extension. This unexpected blue-shift is nicely reproduced by the computed excitation energy of the DE and SE states at TDUDFT level (Figures 4b) or NEVPT2 level (Figure S21 and Table S6) for longer members in the series. A further, blue-shifted DE state is predicted for **TTM-ph-ph-TTM**. This blue shift is readily rationalized by the CR nature of the DE state and by the increasing separation between radical centers. Focusing on the Hubbard dimer model, upon increasing the inter-radical distance the inter-site Coulomb repulsion, $V$, decreases, leading to an increase of the effective on-site Coulomb repulsion $U_{eff}$, a trend which is confirmed by DFT calculations which predict increasing $U_{eff}$ values (11.6 kcal/mol, 26.6 kal/mol and 38.3 kcal/mol) from **TTM-TTM** to **TTM-ph-ph-TTM**. Higher energy excited states of the three investigated diradicals are safely ascribed to localized excitations or, to be more specific, to linear combinations of local excitations centred on radical fragments (**TTM**-related bands in Figure S20), as shown by the fragment orbital analysis and by the TheoDORE analysis.[50]

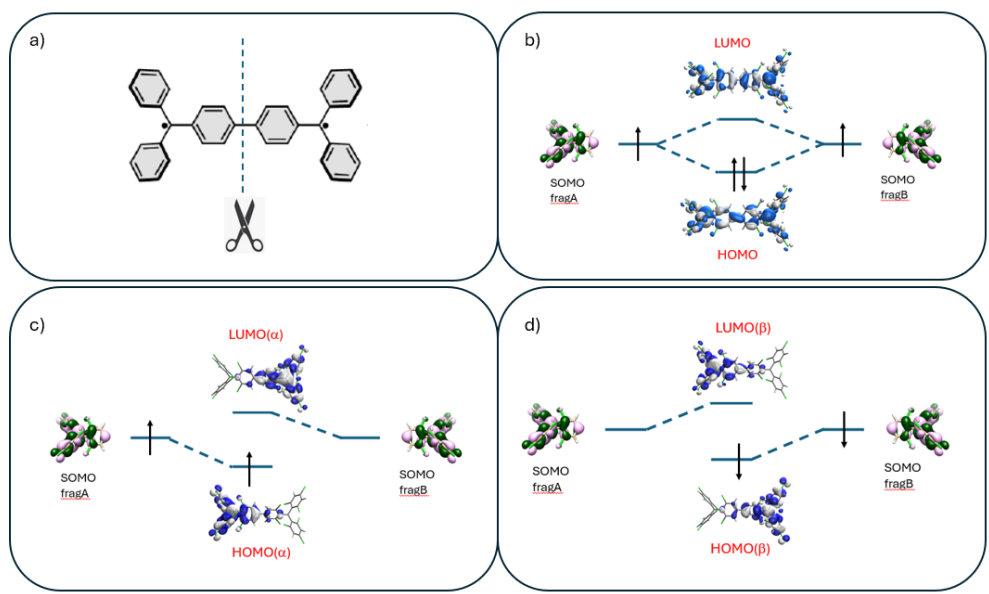

**Figure 3.** a) Schematic representation of the fragment orbital approach and interaction diagrams (in terms of **TTM** fragment SOMOs) for the formation of the MOs of **TTM-TTM**. b) delocalized HOMO and LUMO computed at the optimized UDFT geometry. c) localized HOMO($\alpha$) and LUMO($\alpha$) computed at the optimized UDFT geometry. d) same as c) but for $\beta$ electrons and orbitals.

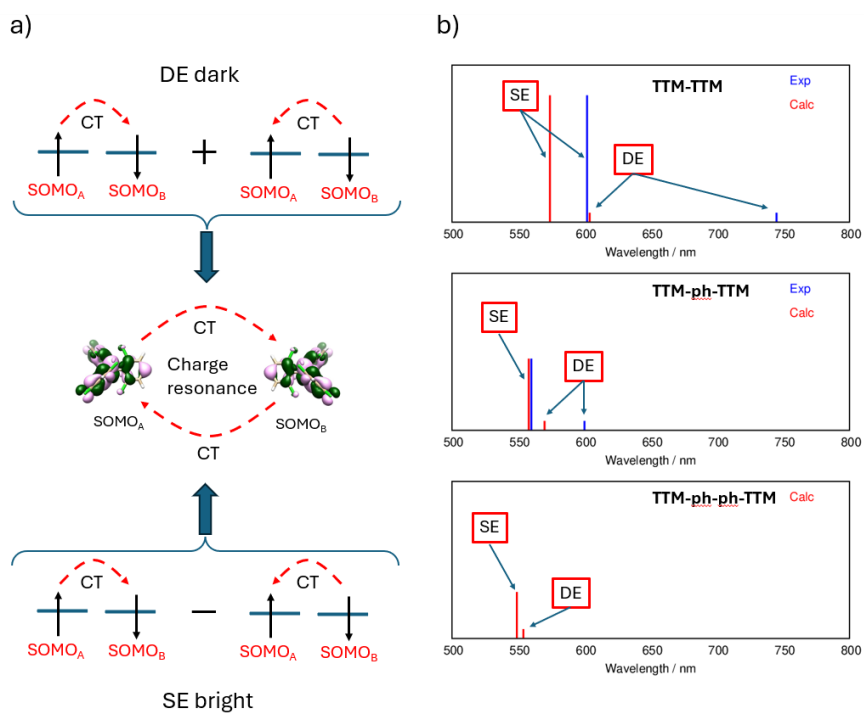

**Figure 4.** a) Schematic representation of the charge resonance (i.e. combination of charge transfer excitations) nature of the bright SE and dark DE (the weakly emitting state) and b) proposed assignment of the transitions observed (blue lines) in absorption for **TTM-TTM** and **TTM-ph-TTM** based on the comparison with predicted transitions (from TDUDFT calculations, red lines). Only computed results are reported for **TTM-ph-ph-TTM**. The intensities of the SE transitions are scaled according to the computed oscillator strengths. For the DE transition an arbitrary low intensity is attributed since the computed oscillator strength (and the observed absorption intensity) is negligible. The overestimation of the DE transition energy for **TTM-TTM** can be reconciled by considering that the computed value corresponds to the vertical excitation while the experimental band corresponds to the 0-0 adiabatic transition. Finally, the onset of the absorption spectrum of **TTM-ph-TTM** at ca. 600 nm[16] is assigned to the DE transition.

Thus, the excited states of **TTM-TTM** and related diradicals featuring a singlet ground state can be grouped in two classes, (Figure S20): high energy states that correlate with the excited states of **TTM** or radical-related, and *new* DE/SE low energy states diradical-related. The crucial point is the relative order of these groups of states. As discussed above, the *new* DE/SE states are the lowest states for **TTM-TTM, TTM-ph-TTM and TTM-ph-ph-TTM**. These are the CR states emerging either from the 2e-2o model or the Hubbard dimer model. However, in both the 2e-2o and the Hubbard models the bright SE state is always located at lower energy than the dark DE state, a result that sharply contrast with the experiment. To regain the correct order of the two states, as experimentally documented for several synthesized diradical molecules[51–55], dynamical electron correlation effects must be properly accounted for, effects which go beyond the 2e-2o or the Hubbard model.

To validate the picture emerging from the calculations, **TTM-TTM** has been synthesized and spectroscopically characterized.

**Synthesis of TTM-TTM**

**TTM-TTM** was synthesized through a novel procedure starting directly from **TTM**. Heating **TTM** powder above its melting point (280 °C), led to the formation of a dark solid (**Scheme 1**). Thin-layer chromatography (TLC) analysis of the product revealed a small amount of a light-blue stain, identified as **TTM-TTM**. Initially, the thermally activated reaction yielded a low amount of the product($Y \approx 5\%$). In fact, once a small amount of **TTM-TTM** is formed, the reaction mixture solidified, inhibiting further reactions. To enhance the reaction kinetics, different amounts of Cu powder were added (see Supplementary Information). Under optimal conditions (20 minutes at 280 °C with 10% Cu), the Ullmann homocoupling of **TTM** resulted in an improved yield of 50%. Prolonged reaction times resulted in the formation of new products, decreasing the reaction yield of **TTM-TTM**. This straightforward synthetic approach leverages the unique reactivity of **TTM**. In this context, when perchloro-triphenyl methyl radical (**PTM**) is heated, the most reactive chlorine atoms, the two ortho- atoms react, resulting in a cyclization to form the perchloro-9-phenylfluorenyl radical (**PPF** in **Scheme 1a**).[56] In contrast, for **TTM**, the most reactive halogens are the para-chlorines, likely due to the reduced dihedral angle of **TTM** compared to **PTM**,[57] and the lower steric hindrance affecting the para-chlorines. This synthetic route resulted more efficient and simpler compared to previously reported pathways.[14,28] Additionally, we do not exclude the possibility that this approach could be extended to the synthesis of more elongated **TTM** oligomers, polymers, or covalent organic frameworks, without the need for the para-functionalization of **TTM**, or its precursor, with more reactive halogens.[58] In this context, the thermal gravimetric analysis (TGA) of **TTM-TTM**, shows similar behaviour to **TTM** but at higher temperature (around 380 °C), suggesting a similar reactivity.

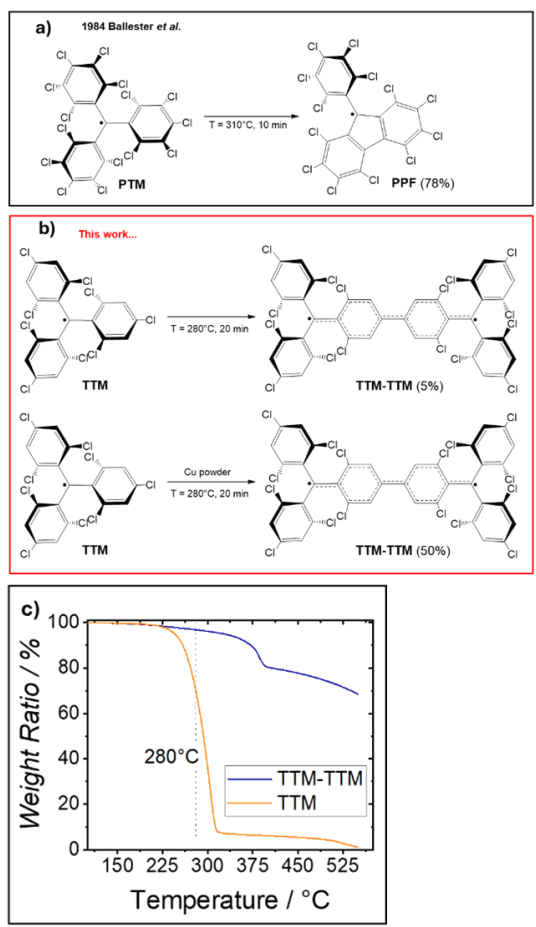

**Scheme 1.** (a) Thermal activated cyclization of **PTM**; (b) Thermal activated dimerization of **TTM** with and without the use of Cu as catalyst; c) TGA spectra of **TTM** and **TTM-TTM**.

**Photophysical properties of TTM-TTM**

The photophysics of **TTM-TTM** was studied in 2-methyltetrahydrofuran (2-meTHF), collecting spectra at variable temperatures between room temperature and 77 K (vitrified solvent). Results are shown in Figure 5. The main absorption band of **TTM-TTM** is observed at ~ 600 nm at room temperature (panel a) with a small red shit at 77 K (panel c); weaker absorption peaks are observed at lower energies. In agreement with the calculations (Figure S20), the absorption bands in the 350-450 nm region can be safely ascribed to excitations localized on the **TTM** fragments, as demonstrated by the comparison between the absorption spectrum of **TTM-TTM** and **TTM** shown in Figure 5a. In this region, the band observed at ~425 nm is attributed to a triplet-triplet transition of **TTM-TTM**, in agreement with the assignment made by von Delius *et al.*[14] Indeed, the band is present at 290 K, when the lowest energy triplet state is thermally populated, but disappears at 77 K.

In the Hubbard model the intensity of the CR transitions can be addressed in the Mulliken approximation, i.e. neglecting all matrix elements of the dipole moment operator but the very large permanent dipole moment $\mu_0$ associated to the charge-separated configurations A$^2$ and B$^2$ ($\mu_0 \approx \pm ed$, where $d$ is the distance between the two radical centers). In this approximation, the DE state has vanishing transition dipole moment, while the transition dipole moment associated to the SE state is proportional to $\mu_0$ with a proportionality constant $x = t/U_{\text{eff}}$. We therefore expect that the intensity of the SE absorption decreases in the series **TTM-TTM**, **TTM-ph-TTM** and **TTM-ph-ph-TTM**. More in detail, we can relate the intensity of the charge-resonance transition to the weight of the zwitterionic configuration in the ground state. For **TTM-TTM** the experimental oscillator strength associated to the absorption band at 600 nm amounts to 0.21, corresponding to a transition dipole moment ~5.2 D. Estimating the distance between the two radical centers as ~10.1 Å, corresponding to a dipole µ₀ ≈ 49 D, the weight of the zwitterionic state in the ground state amounts to $x^2 \approx 0.01$. Quite interestingly, the sizable intensity of this allowed charge-resonance transitions is "borrowed" from the local **TTM-**centred excitations, whose oscillator strength is largely suppressed in the diradical (see Figure 5, panel a). In **TTM-ph-TTM** the increased distance between the radical centers leads to a reduced delocalization, and hence to a less-effective borrowing of intensity of the CR bands from the localized transitions (cf Figure 1).

Emission of **TTM-TTM** occurs in the near-IR, at wavelengths longer than 700 nm, far away from the region where **TTM** emits (see the comparison in Figure 5b). Emission spectra recorded at different temperatures, from 290 K to 77 K, are shown in Figure 5d. The solvent is liquid from 290 to 150 K, but forms a glass at 77 K. Regardless of temperature, luminescence comes from a low-energy dark state (the DE state) and not from the state responsible for the main absorption at 600 nm (the SE state). In liquid solvent, the position, shape and intensity of the emission spectrum are barely dependent on temperature, with the emission maximum located at ~805 nm. In the glassy solvent, the rigidochromic effect is responsible for a blue-shift ($\lambda_{em}^{max} = 765$ nm) and the vibronic structure becomes more resolved, as the thermal disorder is reduced.

The corresponding excitation spectra at different temperatures are dominated by the band at 600 nm, as in absorption (Figure S22), and are scarcely temperature-dependent. Interestingly, in the excitation spectra the band at ~425 nm is absent also at 290 K, supporting again its triplet-triplet nature.

The excitation anisotropy spectrum of **TTM-TTM** in glassy 2-meTHF solvent at 77 K, collected in the whole experimentally accessible spectral window (300-770 nm), is shown in Figure 5c. At this temperature, the rotational motion of the solute is hindered, so that the measured anisotropy corresponds to the so-called fundamental anisotropy, $r_0$, related to the angle $\alpha$ between the absorption and emission transition dipole moments by the relation $r_0 = \frac{2}{5}\left(\frac{3\cos^2\alpha - 1}{2}\right)$. The fundamental anisotropy ranges from 0.4, for perfectly aligned emission and absorption transition dipole moments ($\alpha = 0$) to -0.2 for orthogonal transition dipoles ($\alpha = 90°$). The anisotropy data in Figure 5c therefore give reliable information about the relative polarization of electronic transitions. Anisotropy of **TTM-TTM** is high (~0.34) and constant over the entire 550-770 nm range, indicating that all the electronic transitions occurring at $\lambda > 550$ nm (the transitions from the ground state to the SE and DE states), have the same polarization as the emission, along the charge-transfer axis of the diradical, as confirmed by the calculated transition dipole moments. At higher energies ($\lambda < 550$ nm), instead, the anisotropy assumes different values, indicating that the higher-energy transitions are characterized by different polarizations. Specifically, at 365 nm, in the region where a transition localized on **TTM** occurs, anisotropy approaches -0.13, corresponding to $\alpha \approx 60°$, supporting a different nature of this state (localized **TTM**-related band) compared to the lowest-energy ones (SE and DE), that are instead new states of the diradical.

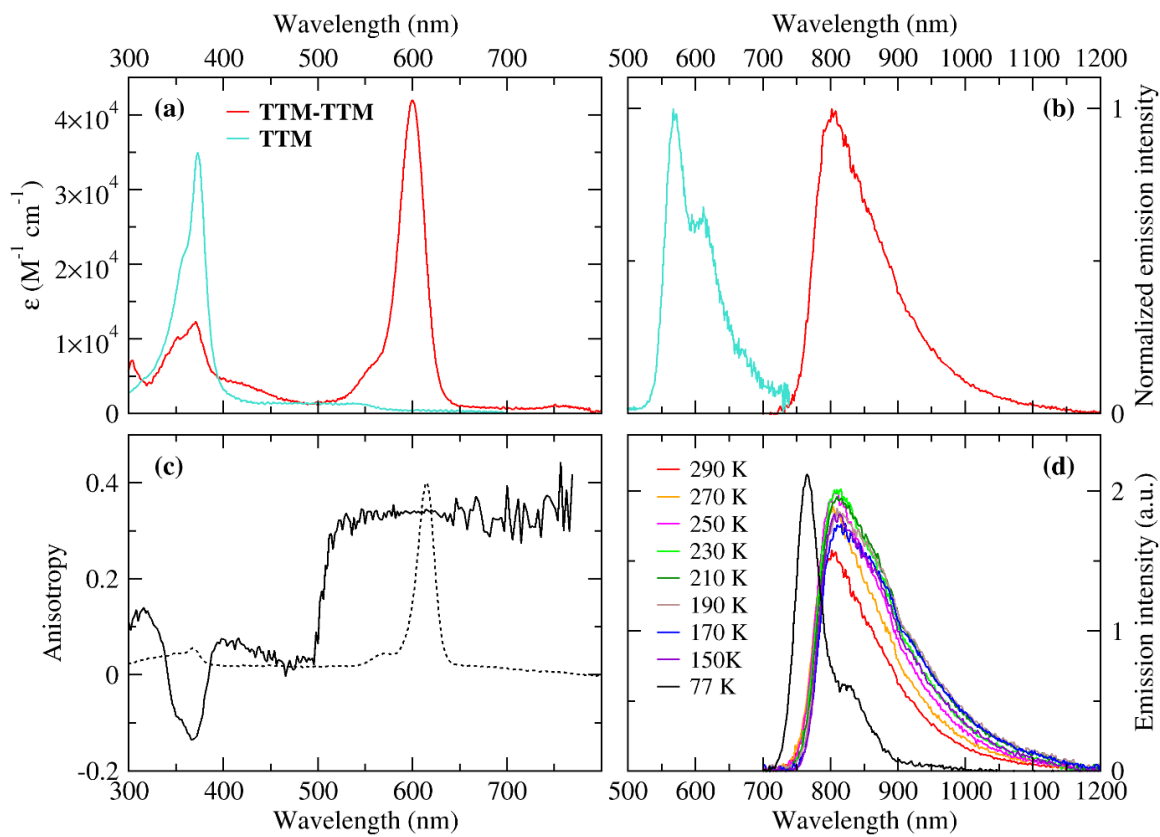

**Figure 5.** (a) Absorption spectrum of **TTM-TTM** in 2-meTHF (red line) and **TTM** in THF (cyan line) at 290 K; (b) Emission spectrum of **TTM-TTM** in 2-meTHF (red line) and **TTM** in THF (cyan line) at 290 K; (c) Excitation anisotropy of **TTM-TTM** (continuous line) and corresponding absorption spectrum (dotted line) in vitrified 2-meTHF solution at 77 K (emission was collected at 800 nm); (d) Emission spectra of **TTM-TTM** in 2-meTHF at different temperatures (excitation wavelength: 600 nm).

## Conclusions

To clarify the relationship between the emission properties of **TTM** radicals and **TTM**-derived diradicals displaying a singlet ground state, we conducted a combined computational and experimental investigation that reveals the nature of the lowest excited states in these diradical species. The lowest-lying excited states of the **TTM**-derived diradicals cannot be related to those of **TTM**: new low-lying excited states emerge, corresponding to charge resonance between the two radical units, the lowest one being a dark state for symmetric diradicals.

This observation aligns with the CR nature of the lowest singlet excited states (DE and SE) as revealed by both the simple 2e-2o and the Hubbard dimer models, thereby offering a unified understanding of the lowest excited states of diradicals. However, electron correlation is essential for accurately predicting the energy order of the two CR states. While the zwitterionic character of the lowest excited state has been previously reported for specific examples of diradicals,[15,59] this work provides a systematic framework to understand the nature of emitting states in singlet ground-state diradicals, emphasizing the dark nature of the lowest excited state. The DE state, previously identified as the lowest excited state in various classes of conjugated diradicals[39–44,51–55] and in **TTH**,[12] also plays a key role in **TTM**-derived diradicals with extended π-conjugated spacers, contributing to their red-shifted luminescence relative to **TTM**. Interestingly, the calculations reveal that, contrary to expectations, the charge resonance character causes the excitation energies of both the SE and DE states to blue shift with increasing π-spacer length, accounting for the observed blue-shifted emission of **TTM-ph-TTM** compared to **TTM-TTM**.

Extensive, variable temperature spectroscopic characterization of **TTM-TTM** supports the theoretical analysis showing that higher energy absorption bands in the 350-450 nm region can be ascribed to excitations localized on the **TTM** fragments while the lowest energy absorption spectrum is due to transitions to the SE and DE states. The data show that luminescence originates from the low-energy dark state (the DE state) and not from the state responsible for the main absorption at 600 nm (the SE state). Furthermore, the excitation anisotropy spectrum of **TTM-TTM** in glassy solvent at 77 K indicates that all the electronic transitions occurring at $\lambda > 550$ nm (the transitions from the ground state to the SE and DE states), have the same polarization as the emission transition, along the charge-transfer axis of the diradical, as confirmed by the calculated transition dipole moment, confirming their CR nature.

We believe that the insights gained from this study highlight that different approaches than those used for **TTM** radical are needed to achieve high PLQY in **TTM**-derived diradicals. According with reported results, a plausible strategy to increase the luminescence efficiency of the lowest excited states of this kind of diradicaloid luminescent species could be to break their symmetric structures. Besides, this work can serve as a guide for designing new diradicals with enhanced luminescence properties.

## ACKNOWLEDGMENT

D.M. gratefully acknowledges the PON-RI Industrial Chemical and Molecular Sciences PhD XXXVII cycle action for funding, an initiative co-founded by European Union through the PON 2022-2025 for Green and Innovation topics (ID grant 634338). M.O., D.B., A.P. and F.N. acknowledge funding from Ministero dell'Università e della Ricerca (MUR) PRIN 2022 project 202253P3YJ MULTIRADICALS4LIGHT: Design, synthesis, and characterization of inert MULTIfunctional diRADICALoidS for organic LIGHT-emitting transistors. A.P. and G.F. acknowledge PNRR MUR project PE0000023-NQSTI. B.B., A.P. and F.T. acknowledge the equipment and framework of the COMP-R Initiatives, funded by the "Departments of Excellence program of the Italian Ministry for University and Research" (MUR, 2023–2027).